\documentclass[reprint,showpacs,showkeys,amsmath,amssymb,aps]{revtex4-1}

\usepackage{graphicx}
\usepackage{dcolumn}
\usepackage{bm}
\usepackage{mathrsfs}

\def\lesim{\lower .7ex\hbox{$\;\stackrel{\textstyle
<}{\sim}\;$}}
\def\gesim{\lower .7ex\hbox{$\;\stackrel{\textstyle
>}{\sim}\;$}}

\begin{document}

\title{Calculations of electric fields for radio detection of Ultra-High Energy particles} 

\author{Daniel Garc\'\i a-Fern\'andez}
\author{Jaime Alvarez-Mu\~niz}
\author{Washington R. Carvalho Jr.}
\affiliation{%
Depto. de F\'\i sica de Part\'\i culas
\& Instituto Galego de F\'\i sica de Altas Enerx\'\i as,
Universidade de Santiago de Compostela, 15782 Santiago
de Compostela, Spain
}%

\author{Andr\'es Romero-Wolf}
\affiliation{%
Jet Propulsion Laboratory, California Institute of Technology, 4800 Oak Grove Drive, 
Pasadena, California 91109, USA
}%

\author{Enrique Zas}
\affiliation{%
Depto. de F\'\i sica de Part\'\i culas
\& Instituto Galego de F\'\i sica de Altas Enerx\'\i as,
Universidad de Santiago de Compostela, 15782 Santiago
de Compostela, Spain
}%

\begin{abstract}
The detection of electromagnetic pulses from high energy showers  
is used as a means to search for Ultra-High Energy cosmic ray and neutrino interactions. 
An approximate formula has been obtained to numerically evaluate the radio 
pulse emitted by a charged particle that instantaneously accelerates, moves 
at constant speed along a straight track and halts again instantaneously. The approximate solution is applied 
to the particle track after dividing it in smaller subintervals. The resulting algorithm (often 
referred to as the ZHS algorithm) is also the basis for most of the simulations of the electric field 
produced in high energy showers in dense media. In this work, the electromagnetic pulses as predicted 
with the ZHS algorithm are compared to those obtained with an exact solution of the electric 
field produced by a charged particle track. The precise conditions that must apply for the algorithm 
to be valid are discussed and its accuracy is addressed. This comparison is also made for electromagnetic 
showers in dense media. The ZHS algorithm is shown to describe Cherenkov radiation and to be valid for 
most situations of interest concerning detectors searching for Ultra-High Energy neutrinos. 
The results of this work are also relevant for the simulation of pulses emitted from air showers.

\end{abstract}

\maketitle

\section{Introduction}

The study of Ultra High Energy Cosmic Rays (UHECR) and neutrinos
(UHE$\nu$s) is currently a high priority in Astroparticle Physics 
with many experimental efforts being dedicated to these two related
areas of research. UHECRs are routinely detected through the Extensive Atmospheric Showers (EAS) 
they produce when interacting in the atmosphere. 
Despite the recent advances in the measurement of the flux of UHECRs
~\cite{HiRes_spectrum,Auger_spectrum}, their primary composition remains unknown~\cite{Auger_Xmax,HiRes_Xmax},
and this is one of the main obstacles to extract precise conclusions on their origin. 
There are strong reasons to believe that 
UHE$\nu$s should be produced in the interactions of UHECRs with the
material surrounding the sources, and/or in their propagation through
the observed Cosmic Microwave Background radiation \cite{Olinto}. 
However their detection has not yet been achieved. Efforts are being made to improve the
experimental situation in both fields. The challenge is
to instrument sufficiently large target volumes to compensate for the low
cross section and low fluxes for UHE$\nu$ detection, and to find measurements 
that provide large aperture at the highest energies and help to constrain 
the primary composition of UHECRs. The radio technique is being
explored in both fields. 

As early as in 1962, G.~Askaryan proposed to detect UHECRs and UHE$\nu$s by observing
the coherent radio pulse from the excess of electrons in a shower developing in a dense, 
dielectric and nonabsorptive to radiowaves medium~\cite{Askaryan62}.
Soon after, pulses were observed in coincidence with air shower arrays~\cite{Jelley,Allan}. 
The emission is coherent in wavelengths which are large compared to
the characteristic size of the electric charge and current
distributions associated to the induced showers. 
The technique has been receiving a lot of attention in the field of 
Astroparticle Physics in the last decade because of the 
relatively low cost of the antennas needed for the detection systems. 
Also, coherence implies that the pulse energy scales with
the square of the primary energy which favours long range detection,
a requirement to achieve large areas and volumes both for UHECR and UHE$\nu$ detection. 
Indeed, quadratic scaling has been confirmed in accelerator experiments~
\cite{Saltzberg_SLAC_sand,Miocinovic_SLAC_sand,Gorham_SLAC_salt,Gorham_SLAC_ice}
leading to very strong pulses associated to UHE energy showers. 

A large number of initiatives have been made or are currently in
development or planning stages. 
In dense media neutrino-induced showers are less than a meter
in width and full coherence is expected to be mantained up to the GHz
range (see for instance \cite{ZHS92}). A variety 
of past and present experiments search for pulses at those frequencies produced by neutrinos in 
ice~\cite{RICE03,ANITA_2009_limits}, the moon
regolith~\cite{Parkes96,GLUElimits,NuMoon,Kalyazin,LUNASKA,RESUN} or salt domes~\cite{Gorham_SLAC_salt}.
The largest and most promising experiments are in planning stages
looking at ice~\cite{ARA,ARIANNA}. 
Electromagnetic pulses emitted in the MHz-GHz frequency range by EAS 
induced by UHECR are being measured in the hope of using this complementary
information to constrain its composition and/or to develop new cost
effective detection systems~\cite{AERA,CODALEMA}. In addition the
ANITA balloon flown antennas, initially devised to search for neutrinos, has 
recorded coherent pulses up to the GHz range~\cite{ANITA_UHECR} consistent with EAS 
induced by UHECRs and there are plans for future developments~\cite{EVA}. 

The success of the radio technique requires an accurate and computationally efficient calculation 
of the radio emission properties of UHE showers. It is necessary to
perform this in an efficient way since the number of particles in a shower
at EeV energies is $\geq 10^9$.  
Simulation techniques for evaluating pulses in dense media have been used for more than 20 
years~\cite{ZHS91,ZHS92,alz97,alz98,alvz99,alvz00,razzaque01,almvz03,
razzaque04,McKay_radio,alvz06,aljpz09,ARZ10,ZHAireS_ice,ARZ11}. 
The shower is simulated to obtain, for every particle track, the information  
needed to calculate its contribution to the electric field.
The formula used for this purpose 
stems from an approximate solution of Maxwell's equations in the Fraunhofer limit~\cite{Allan}
adapted for simulation purposes in~\cite{ZHS92}. 
The electric field due to a short charged particle track, assumed to be travelling at
a constant speed, can be approximated by two terms which 
correspond to the start and end of the track. The resulting electric field depends 
on the particle speed and the angle between the line of sight from the track to the observer
~\cite{ZHS92,ARZ10}. To calculate the emission in a shower,
tracks are chosen so that the particle velocity can be approximated to
be constant, and all track contributions are added taking into account 
interference effects. 

Although the approximate formula is sufficient for many practical
applications, its range of validity is limited in frequency and
position of the observer with respect to the track. 
It is possible to extend its range of validity by subdividing each track 
in sub-intervals. The resulting algorithm (often referred to as ``ZHS algorithm") 
is easy to implement and fast enough for the simulation of particle showers. 
This is convenient since it has allowed the simulation of pulses in the Fresnel region for neutrino
detection~\cite{alz98} and in measurements of EAS~\cite{ZHAireS_air}. 
Although the range of applicability of the algorithm is enhanced
when used in this way, 
it is not obvious that the sum of the subcontributions correclty accounts
for all the radiation in regions close to the emission source. 

The object of this article is to study the conditions for the ZHS algorithm 
to be valid and to establish its accuracy. 
For this purpose we obtain in Section II an exact solution to the 
problem of a charged particle instantaneously accelerating to a constant speed
and stopping abruptly after a discrete time interval \cite{Tamm}. This 
allows us to establish the precise conditions necessary to turn this solution 
into the basic formula of the ZHS algorithm. In Section III we compare the exact solutions 
for an infinite and a finite track to the result of the ZHS algorithm, and stress the 
compatible interpretation of the radiation regime in terms of Cherenkov radiation for both cases. 
In Section IV we compare the exact solution of the single track problem
in nearby regions with the results of applying the ZHS algorithm with
track subdivisions to test its validity and accuracy. Section V is devoted to
discussing the ZHS algorithm in relation to other approximations
made to calculate the pulses emitted from showers. Section VI
presents the summary and conclusions of our work.

\section{Electromagnetic field of a single charged particle track}

Let us assume an electron ejected from an atom at time $t=t_1$ that travels at constant speed $\mathbf{v}$ 
through a medium along a finite track until it is absorbed by another atom at time $t_2$. Neglecting 
the movement of the atoms, we can model the electric current associated to the electron as,
\begin{equation}
\mathbf{J}(\mathbf{x},t) = -e \mathbf{v} \delta^{(3)}(\mathbf{x}-\mathbf{x_0}-\mathbf{v}t)
~\Theta(t-t_1) \Theta(t_2-t)
\label{eq:current}
\end{equation}
where $e=\vert e \vert$ is the charge of a positron, 
$\mathbf{x}(t)$ is its position and $\mathbf{x_0}$ an arbitrary reference position.
The step $\Theta$-functions account for the fact that the electron only moves 
in the time interval ($t_1$, $t_2$).  

In a dielectric medium with permitivity $\epsilon$ and magnetic susceptibility $\mu$,   
Maxwell's equations for the vector potential in the frequency domain $\mathbf{A}(\mathbf{x},\omega)$
can be written as~\cite{Jackson}:
\begin{equation}
\begin{split}
&
\nabla^2\mathbf{A}(\mathbf{x},\omega)+\mu\epsilon\omega^2\mathbf{A}(\mathbf{x},\omega) - 
\\
&
\nabla[\nabla\cdot\mathbf{A}(\mathbf{x},\omega)-i\epsilon\mu\omega~\phi(\mathbf{x},\omega)] = -\mu\mathbf{J}(\mathbf{x},\omega),
\end{split}
\label{eq:Avec}
\end{equation}
where $\phi(\mathbf{x},\omega)$ is the Fourier transform of the scalar
potential and we use the following convention for the Fourier-transform of a function $f(t)$:
$f(\omega)=\int_{-\infty}^{\infty} \mathrm{d}t \ e^{i\omega t}f(t)$. 
In principle $\epsilon$ and $\mu$ can depend on frequency and our results below
would be equally valid, but we drop the explicit
dependence of $\epsilon$ and $\mu$ with $\omega$ for simplicity. 

We use the Lorenz gauge condition which implies that Eq.~(\ref{eq:Avec}) for the vector potential becomes:
\begin{equation}
\nabla^2\mathbf{A} + k^2\mathbf{A} = -\mu\mathbf{J},
\label{eq:helmholtz}
\end{equation}
with $k=n\omega/c$ and $n$ the refractive index of the medium.
The Lorenz gauge condition in the frequency domain~\cite{Jackson}: 
\begin{equation}
\nabla\cdot\mathbf{A}(\mathbf{x},\omega) = i\epsilon\mu\omega~\phi(\mathbf{x},\omega),
\label{eq:gauge}
\end{equation}
implies that the scalar potential for non-zero frequencies is entirely
determined by the divergence of the vector and only the vector
potential is needed to calculate the field. 

\subsection{Exact solution}
\label{sec:exact}

The solution of the Helmholtz equation in Eq.~(\ref{eq:helmholtz}) is standard physics and 
can be obtained, using Green's method, as an integral over source positions~\cite{Jackson}:
\begin{equation}
\mathbf{A}(\mathbf{x},\omega) = \frac{\mu}{4\pi}\int\mathrm{d}^3 \mathbf{x'}
\frac{e^{ik|\mathbf{x}-\mathbf{x'}|}}{|\mathbf{x}-\mathbf{x'}|} \mathbf{J}(\mathbf{x'},\omega)
\label{eq:Avec_w}
\end{equation}
where from now on, $\mathbf{x}$ will denote the position of the observer and $\mathbf{x'}$ 
the position of the source.

For simplicity we rewrite the current in Eq.~(\ref{eq:current}) as:
\begin{equation}
\mathbf{J}(\mathbf{x},t) = q v~Z(t)~P(\mathbf{x},t)~\hat z 
\end{equation}
with $q=-e$; $Z(t)=\Theta(t-t_1)\Theta(t_2-t)$ and
$P(\mathbf{x,t})=\delta^{(3)}(\mathbf{x}-vt\hat z - z_0 \hat z)$, 
where we assume without loss of generality that the electron 
travels along the $z$-axis which is parallel to $\hat z$. 
The equations that follow below are clearly valid for  
any continuous and differentiable functions $Z(t)$ and
$P(\mathbf{x},t)$. They are also valid for the case considered in Eq.~(\ref{eq:current}) which can
be obtained as a limiting case of suitable continuous and
differentiable functions.

Transforming this definition of the current to the frequency domain and 
substituting into Eq.~(\ref{eq:Avec_w}) for the vector potential we obtain:
\begin{equation}
\label{eq:vectorpotential}
\mathbf{A}(\mathbf{x},\omega) = \frac{\mu}{4\pi}~qv~\hat z~
\int\mathrm{d}^3\mathbf{x'}\mathrm{d}t'~
e^{i\omega t'}\frac{e^{ik|\mathbf{x}-\mathbf{x'}|}}{|\mathbf{x}-\mathbf{x'}|}
Z(t') P(\mathbf{x'}, t')
\end{equation}

In the Lorenz gauge, the only non-zero component of the vector potential for a charge moving along $z$ 
is the $z$-component. 
The scalar potential $\phi$ is obtained through Eq.~(\ref{eq:gauge}). 
For this purpose we need to obtain $\nabla\cdot\mathbf{A}$:
\begin{eqnarray}
\label{eq:divergence}
\nabla\cdot\mathbf{A}(\mathbf{x},\omega) = \frac{\partial A_z}{\partial z} = \frac{\mu}{4\pi}~qv~
\int\mathrm{d}^3\mathbf{x'}\mathrm{d}t'~Z(t')P(\mathbf{x'},t') \nonumber \\
e^{i\omega t'}~\frac{e^{ik|\mathbf{x}-\mathbf{x'}|}}{|\mathbf{x}-\mathbf{x'}|}
\left[ ik - \frac{1}{|\mathbf{x} - \mathbf{x'}|} \right]
\frac{(z-z')}{|\mathbf{x} - \mathbf{x'}|} \nonumber
\end{eqnarray}

The electric field $\mathbf{E}(\mathbf{x},\omega)$ can be obtained from the scalar and vector potentials
\cite{Jackson} as:
\begin{equation}
\mathbf{E}(\mathbf{x},\omega) = -\nabla \phi(\mathbf{x},\omega) + i\omega\mathbf{A}(\mathbf{x},\omega)
\end{equation}

Due to the cylindrical symmetry of a charged particle track we calculate only
the radial ($E_\rho$) and $z$ ($E_z$) components of the field. 
Since $\mathbf{A}$ only has $z$-component, then the radial component of the field 
is given by:

\begin{equation}
E_\rho(\mathbf{x},\omega) = -\frac{\partial\phi(\mathbf{x},\omega)}{\partial\rho} = 
-\partial_\rho \frac{\nabla\cdot\mathbf{A}(\mathbf{x},\omega)}{i\mu\epsilon\omega}
\label{field_rho}
\end{equation}
where $\rho=\sqrt{x^2+y^2}$ is the radial coordinate, and $\partial_\rho$ denotes $\partial/\partial\rho$.

The $z-$component of the field is given by:
\begin{eqnarray}
E_z(\mathbf{x},\omega) = -\partial_z\phi(\mathbf{x},\omega) + i\omega A_z(\mathbf{x},\omega)  \nonumber \\
= -\partial_z \frac{\nabla\cdot\mathbf{A}(\mathbf{x},\omega)}{i\omega\mu\epsilon} + i\omega A_z(\mathbf{x},\omega)
\label{field_z}
\end{eqnarray}
with $\partial_z$ denoting $\partial/\partial z$.

Performing the derivatives in Eqs.~(\ref{field_rho}) and (\ref{field_z}) we obtain:
\begin{equation}
\begin{split}
&
E_\rho(\mathbf{x},\omega) = 
i~\frac{qv}{\omega}~\frac{1}{4\pi \epsilon}
\int_{t_1}^{t_2}\mathrm{d}t'~ e^{i\omega t'}~\frac{e^{ikr}}{r^3} \times
\\
&
\rho \times (z-z_0-vt') \times \left[b\left(b-\frac{1}{r}\right) + \frac{1}{r^2} \right]
\end{split}
\label{eq:rhofield}
\end{equation}
and,

\begin{equation}
\begin{split}
&
E_z(\mathbf{x},\omega) = 
i~\frac{qv}{\omega}~\frac{1}{4\pi \epsilon}~\int_{t_1}^{t_2} \mathrm{d} t'~ e^{i\omega t'} \frac{e^{ikr}}{r^2} ~\times
\\
& 
\Bigg[b^2 \frac{(z-z_0-vt')^2}{r} ~+~
\frac{(z-z_0-vt')^2}{r^3} ~-~ 
\\
&
b \left( \frac{(z-z_0-vt')^2}{r^2} -1 \right) \Bigg] ~+~ 
\\
&
i\omega~\frac{\mu}{4\pi}~qv~\int_{t_1}^{t_2} 
\mathrm{d}t'~e^{i\omega t'}~\frac{e^{ikr}}{r}
\end{split}
\label{eq:zfield}
\end{equation}
where $r=r(t')$ and $b=b(t')$ are both functions of the source time $t'$ which is defined as, 
\begin{equation}
r(t') = |\mathbf{x}-\mathbf{x'}| = \sqrt{\rho^2+(z-z_0-vt')^2}
\end{equation}
and
\begin{equation}
b(t') = ik - \frac{1}{r(t')}
\end{equation}

Eqs.~(\ref{eq:rhofield}) and (\ref{eq:zfield}) provide an exact solution for the electric field of a finite track.
In general they do not have an analytical form and a numerical integration needs to be performed to 
obtain the field. For the calculations in this paper, we have divided the integration interval
and applied Simpson's rule, increasing the number of divisions until the integral converged.
Under certain conditions, however, we can give
analytical approximations for relevant physical situations.

In the following we show that the basic expression used in the ZHS
algorithm is a particular case of Eqs.~(\ref{eq:rhofield}) and (\ref{eq:zfield}) 
under certain approximations. 

\subsection{The ZHS expression}
\label{sec:ZHS}

A simple expression for the approximate calculation of the electric field 
from a single charged particle track moving at constant speed was
found in \cite{ZHS92}.  In this section we
derive the expression used for the ZHS algorithm from the exact
solution and compare the electric field as obtained in both the exact
calculation and with the ZHS expression. This allows us to 
establish under which circumstances the formula gives a good account of 
the electric field.  

The ZHS algorithm can be obtained from Eqs.~(\ref{eq:rhofield}) and (\ref{eq:zfield}) 
if the following set of conditions are fulfilled:

\begin{enumerate}

\item 
The observer is in the ``far field" zone i.e. 
\begin{equation}
k r \gg 1
\label{eq:cond1}
\end{equation}

\item 
The Fraunhofer approximation holds. This can be stated as a condition
for the phase factor to be approximated as: 
\begin{equation}
kr = k\vert \mathbf{x}-\mathbf{x'} \vert \approx k[R - v(t-t_0)\cos\theta] 
\label{eq:fraunhofer}
\end{equation}
where $R$ is the distance
from the observation point to a reference point along the track where the particle 
is located at a reference time $t_0$, and $\theta$
is the angle between the particle track and the direction from the reference point  
to the observer. This approximation holds provided the parameter $\eta
\ll 1$ with $\eta$ defined as: 
\begin{equation}
\eta(t) = \frac{k [v(t-t_{\rm 0})]^2}{R}\sin^2\theta, 
\label{eq:cond3}
\end{equation}
This condition should be fulfilled at any time $t$ from $t_1$ to $t_2$.
A more commonly used and nearly equivalent form of this condition~\cite{buniy02} is:   
\begin{equation}
\eta' = \frac{k L^2}{R}\sin^2\theta \ll 1, 
\label{eq:cond3bis}
\end{equation}
where $L=v(t_2-t_1)$ is the length of the track. 
This condition is necessary to ensure that the second and higher order terms for the phases 
$i(\omega t + kr)$ in Eqs.~(\ref{eq:rhofield}) and (\ref{eq:zfield}) have no relevance, 
even when the sum of the leading and first order terms in the Taylor expansion of the phases 
is zero, as it occurs for observation at the Cherenkov angle \cite{Afanasiev_book} defined 
as $\cos\theta_C=1/\beta n$ with $\beta=v/c$. 

\item 
Finally the distance to the observer appearing in the denominators of several terms 
of Eqs.~(\ref{eq:rhofield}) and (\ref{eq:zfield}) must be approximated as: 
\begin{equation}
\frac{1}{r(t)} \approx \frac{1}{R}
\label{eq:cond4}
\end{equation}
over the length $L$ of the track, where $R$ is the distance to a reference point along 
the track (in the ZHS algorithm the mid-point of the track is selected). 
The error when making this approximation is of order $L/R$. 

\end{enumerate}

In particular, the condition in Eq.~(\ref{eq:cond1}) implies:
\begin{equation}
b(t') \approx ik ~; \qquad  b - \frac{1}{r}\approx b ~; \qquad  b^2 + \frac{1}{r^2} \approx b^2 
\label{eq:b}
\end{equation}

With these approximations 
the radial component of the field in Eq.~(\ref{eq:rhofield}) becomes:
\begin{equation}
\begin{split}
&
E_\rho \approx i \frac{qv}{\omega}~\frac{1}{4\pi \epsilon}~\frac{e^{ikR}}{R}~(ik)^2~
\sin\theta\cos\theta~ e^{i  \mathbf{k}\cdot\mathbf{v} t_0}~
\\
&
\int_{t_1}^{t_2}\mathrm{d}t'~ e^{i(\omega-\mathbf{k}\cdot\mathbf{v})t'}
\end{split}
\label{eq:rhoZHS}
\end{equation}
where we have used:
\begin{equation}
\frac{\rho}{r} \approx \frac{\rho}{R} = \sin\theta, 
\end{equation}
and 
\begin{equation}
\frac{z-z_0-vt}{r} \approx \frac{z-z_0-v t_0}{R} = \cos\theta.
\end{equation}

Eq.~(\ref{eq:rhoZHS}) can be easily integrated yielding:
\begin{equation}
\label{eq:rhofield_ZHS}
\begin{split}
&
E_\rho = -iq~\omega ~\frac{\mu}{4\pi}~v\sin\theta\cos\theta~ \frac{e^{ikR}}{R}~
\\
&
e^{i\mathbf{k}\cdot\mathbf{v} t_0}~\left[ \frac{e^{i(\omega-\mathbf{k}\cdot\mathbf{v})t_2}
-e^{i(\omega-\mathbf{k}\cdot\mathbf{v})t_1}}{i(\omega-\mathbf{k}\cdot\mathbf{v})}
\right] .
\end{split}
\end{equation}
If we make $t_0=t_1$ this becomes the expression for the radial
field as used in the ZHS algorithm \cite{ZHS92}
except for a factor 2 due to the Fourier transform convention used in \cite{ZHS92}.

Similarly applying the approximations in Eqs.~(\ref{eq:fraunhofer}), (\ref{eq:cond4}) and (\ref{eq:b}) 
to Eq.~(\ref{eq:zfield}) for the $z$-component of the field, and using that $k R \gg 1 \Rightarrow k^2 \gg k/R$,
it is straightforward to show that:
\begin{equation}
\begin{split}
&
E_z \approx -i\omega^2~\frac{q v}{\omega}\frac{\mu}{4\pi}~\frac{e^{ikR}}{R}~
e^{i\mathbf{k}\cdot\mathbf{v}~t_0} 
\cos^2\theta~\int_{t_1}^{t_2} \mathrm{d} t \ e^{i(\omega  - \mathbf{k}\cdot\mathbf{v})t} 
\\
&
+ iqv~\omega ~\frac{\mu}{4\pi}~\frac{e^{ikR}}{R}~
e^{i\mathbf{k}\cdot\mathbf{v}t_0}~\int_{t_1}^{t_2}\mathrm{d}t \
e^{i(\omega  - \mathbf{k}\cdot\mathbf{v})t}~,
\end{split}
\label{eq:zZHS}
\end{equation}
which can be cast as: 
\begin{equation}
\label{eq:zintzhs}
E_z = iq~\omega ~\frac{\mu}{4\pi}~v\sin^2\theta~\frac{e^{ikR}}{R}~e^{i\mathbf{k}\cdot\mathbf{v} t_0}
\int_{t_1}^{t_2} \mathrm{d} t \ e^{i(\omega - \mathbf{k}\cdot\mathbf{v}) t}
\end{equation}

Performing the integral, and taking $t_0=t_1$, the ZHS formula is recovered:

\begin{equation}
\label{eq:zfield_ZHS}
\begin{split}
&
E_z = iq~\omega \frac{\mu}{4\pi}~v\sin^2\theta~\frac{e^{ikR}}{R}~
\\
&
e^{i\mathbf{k}\cdot\mathbf{v} t_1}~\left[ \frac{e^{i(\omega-\mathbf{k}\cdot\mathbf{v})t_2}~
-e^{i(\omega-\mathbf{k}\cdot\mathbf{v})t_1}}{i(\omega-\mathbf{k}\cdot\mathbf{v})} \right]~.
\end{split}
\end{equation}

\subsection{The ZHS algorithm}
\label{sec:ZHSAlgorithm}

To calculate the electric field of the pulse emitted from a current distribution, such as
that produced in a high energy shower, the ZHS algorithm uses 
the ZHS expressions in Eqs.~(\ref{eq:rhofield_ZHS}) and (\ref{eq:zfield_ZHS}) to
calculate the emission from all the charged particle tracks. 
The final result is obtained adding up all the contributions. 
The value of $R$ used for the phase factor in each particle track is
the distance from the first point of the track to the observation point. 
This definition is consistent with the convention to account for the
phase change between emission arising from the start and end points 
of the track. The actual value of $R$ used for the denominator is the
distance between the midpoint of the track and the observer which is 
for all practical purposes the same value when the approximation is 
valid. The algorithm used in alternative simulation programs is 
similar but different in these technical details~\cite{endpoints}.

Naturally it is possible to divide any charged particle track into
arbitrarily small subtracks in order to make the computation more
accurate. This will extend the range of validity of the
approximation. It is interesting to note that it makes
absolutely no difference to subdivide the track of a uniformly moving
charge when observed in the Fraunhofer limit because the term 
associated to the end of one subtrack cancels the term
associated to the beginning of the next subtrack. This is because in
this limit the differences in $R$ between adjacent subtracks are
arbitrarily small. However in practical situations this cancelation is
not exact if $R$ is allowed to change from a track to the next one according to geometry. 
In the original ZHS program~\cite{ZHS91} the tracks were not subdivided 
but soon it was realized that a more
accurate result was obtained by subdividing all tracks at every point
there was a discrete interaction in the simulation
program~\cite{ZHS95,alvz00}. This was the main modification that was ever made to the
original ZHS Monte Carlo \cite{ZHS91,ZHS92} and has been effective since then. 
With this subdivision the distribution of tracks for a shower in ice has a peak 
at about $L\sim1~$mm. 

\section{Cherenkov radiation}
\label{sec:Cherenkov}

The exact electric field for a charged particle track derived in Eqs.~(\ref{eq:rhofield}) and (\ref{eq:zfield}) 
must account for every single feature of the electric field, since no approximations have been made. 
In particular, it must reproduce Cherenkov radiation which is the only
radiation emitted by a charged particle moving at constant speed, 
when $v>c/n$ in the limit of an infinite track. An analytical solution for the electric field produced
by such particle can be obtained and it is given in \cite{Afanasiev_book}(chapter 4). 
The radial $\rho$ and $z-$components of the fields 
given in \cite{Afanasiev_book} converted to SI units, and using the Fourier transform convention adopted 
in this work, can be written as:
\begin{equation}
\label{eq:rhofield_inf}
E_\rho(\rho,z,\omega) = \frac{q}{2\pi\epsilon v} e^{i\frac{\omega}{v}z}u K_1(u\rho)
\end{equation}
\begin{equation}
\label{eq:zfield_inf}
E_z(\rho,z,\omega) = \frac{i\omega\mu q}{2\pi}e^{i\frac{\omega}{v}z} 
\left(1 - \frac{1}{\mu\epsilon v^2}\right) K_0(u\rho)~,
\end{equation}
where $\rho$ is the radial coordinate, $K_0$ and $K_1$ are the modified Bessel functions
of the second kind, and $u=u(\omega)$ is a function that can take two different values 
depending on the magnitude of the particle speed,  
$v<c/n$ or $v>c/n$ (subluminal or superluminal regime). 
In our convention:
\begin{equation}
v < \frac{c}{n} \Rightarrow u(\omega) = \frac{\omega}{v}\left| \sqrt{1 - n^2\beta^2} \right|
\end{equation}
\begin{equation}
v > \frac{c}{n} \Rightarrow u(\omega) = -i \frac{\omega}{v}\left| \sqrt{n^2\beta^2-1} \right| .
\end{equation}
If the particle travels below the speed of light in the medium, the argument $u(\omega)$ of the Bessel
functions is real and the particle does not radiate as shown in \cite{Afanasiev_book}. 
On the contrary, if the speed of the particle is larger than the speed
of light, then $u(\omega)$ is imaginary and the particle radiates. 
The latter case corresponds to pure Cherenkov radiation \cite{Afanasiev_book}.

With the help of the asymptotic forms for the Bessel functions, we can obtain the limits
of Eqs.~(\ref{eq:rhofield_inf}) and (\ref{eq:zfield_inf}) when $|u\rho| \ll 1$ i.e. for small 
distances to the track compared to the radiation wavelength. Conversely we can also obtain 
the limits when $|u\rho| \gg 1$ i.e. for large distances compared to the wavelength.
If  $|u\rho| \ll 1$ the $K_1$ Bessel function
dominates over the $K_0$ and only the radial component of the field $E_\rho$ matters.
In this case,
\begin{equation}
\label{eq:limit_urho_small}
\lim_{|u\rho|\rightarrow 0} |E_\rho| = \frac{|q|}{2\pi\epsilon v} \frac{1}{\rho}
\end{equation}
and a $1/\rho$ dependence with distances is obtained, as well as no dependence with frequency.

If $|u\rho| \gg 1$ and keeping in mind that when $v>c/n$ the argument $u$ is imaginary  
the fields can be written as:
\begin{equation}
\label{eq:limit_rho_urho_large}
\lim_{u\rho \rightarrow \pm i \infty} |E_\rho| = \frac{|q|}{2\pi\epsilon v} \sqrt{\frac{\pi}{2}}
\left |(n^2\beta^2-1)^{1/4} \sqrt{\frac{\omega}{v\rho}} \right |
\end{equation}
\begin{eqnarray}
\label{eq:limit_z_urho_large}
\lim_{u\rho \rightarrow \pm i \infty} |E_z| = \frac{\mu |q|}{2\pi}\left ( 1 - \frac{1}
{\mu\epsilon v^2}\right ) \sqrt{\frac{\pi}{2}} \nonumber \\
\left | \frac{1}{(n^2\beta^2-1)^{1/4}}
\sqrt{\frac{v\omega}{\rho}} \right |
\end{eqnarray}
In this case the field is proportional to $\sqrt{\omega/\rho}$.
This is in agreement with \cite{buniy02} where the same behavior is deduced 
using simple arguments of energy conservation through a cylindrical surface 
surrounding the track. 

\begin{figure}[htbp]
\begin{center}
{
\scalebox{0.35}{\includegraphics[angle=-90]{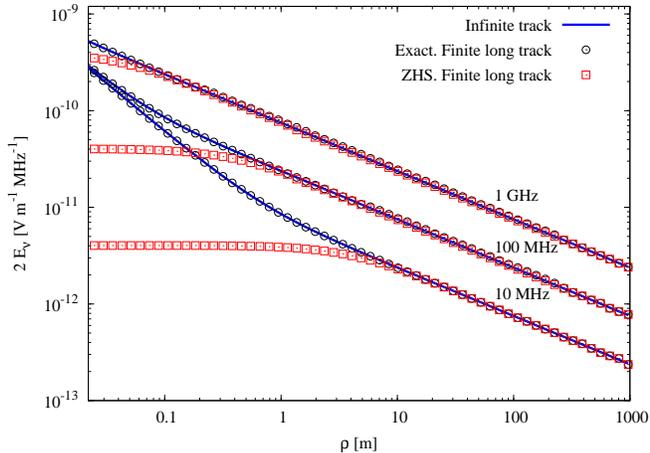}}
}
\caption{
Fourier components of the electric field modulus as a function of distance to
the particle track for an infinite track as obtained from Eqs.~(\ref{eq:rhofield_inf}) 
and (\ref{eq:zfield_inf}) (solid line), and for a track of length $L = 1200$ m as obtained
with the exact formulas derived in this work (Eqs.~(\ref{eq:rhofield}) and (\ref{eq:zfield})) (open circles) 
and with the ZHS algorithm (Eqs.~(\ref{eq:rhofield_ZHS}) and (\ref{eq:zfield_ZHS})) (open squares).
From top to bottom, the observation frequencies are $1$ GHz, $100$ MHz and $10$ MHz.
The $1/\rho$ and $1/\sqrt{\rho}$ regimes are apparent.
}
\label{fig:inf_vs_finzhs}
\end{center}
\end{figure}

In Fig.~\ref{fig:inf_vs_finzhs} the Fourier components of the modulus of the electric field
for an infinite track as obtained from Eqs.~(\ref{eq:rhofield_inf}) and (\ref{eq:zfield_inf})  
are shown, for different frequencies, as a function of $\rho$, the radial distance to the track.
The particle speed is $v\simeq c > c/n$ travelling in homogeneous ice with 
refractive index $n=1.78$. Under these circumstances the quantity $|u\rho|$ can be approximated as:
\begin{equation}
\label{eq:u}
|u\rho| \approx 3 \left(\frac{\nu}{100~{\rm MHz}}\right)~\left(\frac{\rho}{1~{\rm m}}\right)
\end{equation} 

At large distances to the track when $|u\rho| \gg 1$ the fields shown in Fig.~\ref{fig:inf_vs_finzhs} 
scale with distance as $1/\sqrt{\rho}$ and with frequency as $\sqrt{\omega}$, 
in agreement with the aysmptotic field components 
in Eqs.~(\ref{eq:limit_rho_urho_large}) and
(\ref{eq:limit_z_urho_large}). 
This behavior takes place when $\rho > 0.1$, 1 and 10 m for frequencies 
$\nu=1$ GHz, 100 MHz and 10 MHz respectively, in agreement with
Eq.~(\ref{eq:u}), as can be seen in Fig.~\ref{fig:inf_vs_finzhs}. 

As the distance to the track decreases 
and the condition $|u\rho| \ll 1$ starts to be valid, the field behaves as $1/\rho$ and does not depend 
on frequency as expected from Eq.~(\ref{eq:limit_urho_small}). As can be clearly seen in Fig.~\ref{fig:inf_vs_finzhs}, 
the transition from the $1/\rho$ behaviour to $\sqrt{\omega/\rho}$ occurs
at a distance that depends on frequency because
$|u\rho|$ involves the frequency (Eq.~\ref{eq:u}). 
For instance at distances $\rho < 0.01$ m the condition
$|u\rho| \ll 1$ applies for both $\nu=100$ and 10 MHz. The Fourier
component of the field scales with $1/\rho$ and has the same value for the two
frequencies as seen in Fig.~\ref{fig:inf_vs_finzhs}, while this is
not the case for a frequency of $\nu=1$ GHz. 

\begin{figure}[htbp]
\begin{center}
{
\scalebox{0.45}{\includegraphics[angle=0]{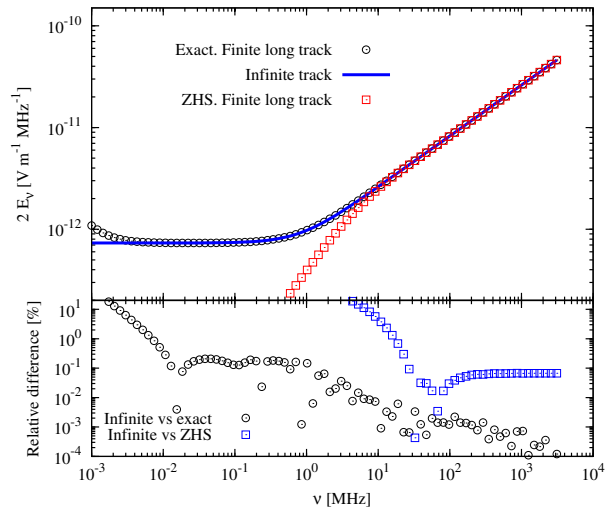}}
}
\caption{
Top panel: Fourier components of the electric field modulus as a function of frequency for an infinite track
as obtained from Eqs.~(\ref{eq:rhofield_inf}) and (\ref{eq:zfield_inf}) (solid line), and for a track of length 
$L = 1200$ m as calculated in this work from Eqs.~(\ref{eq:rhofield}) and (\ref{eq:zfield}) (open circles).
The observer is placed at a lateral distance to the infinite track $\rho = 8.27$ m.
Also shown is the modulus of the field for the same finite track ($L=1.2$ km) as obtained 
with the ZHS algorithm, Eqs.~(\ref{eq:rhofield_ZHS}) and (\ref{eq:zfield_ZHS}) (open squares).
At high frequencies, the field behaves with $\sqrt{\omega}$ (see text for explanations).
Bottom panel: Relative difference (in $\%$) between the solution for an infinite track
and the exact solution for a finite long track (open circles) and between the solution for an infinite track 
and that obtained with the ZHS algorithm (open squares).
} 
\label{fig:inf_vs_fin}
\end{center}
\end{figure}

In Fig.~\ref{fig:inf_vs_fin} 
the modulus of the field for a charged particle in an infinite track - as obtained from Eqs.~(\ref{eq:rhofield_inf}) 
and (\ref{eq:zfield_inf}) - is shown as a function of frequency for an observer at a fixed radial distance.
At large enough frequencies so that the condition $|u\rho|\gg 1$ applies, the field scales as $\sqrt{\omega}$
as expected from Eqs.~(\ref{eq:limit_rho_urho_large}) and (\ref{eq:limit_z_urho_large}), while it is constant with
frequency for small enough frequencies so that $|u\rho|\ll 1$ as predicted from the asymptotic Eq.~(\ref{eq:limit_urho_small}).
More quantitatively, since the observer in Fig.~\ref{fig:inf_vs_fin} is located at $\rho\sim 10$ m,
the field should behave as $\sqrt{\omega}$ for $\nu \gesim 10$ MHz (applying Eq.~(\ref{eq:u})). This is
approximately the case as can be seen in Fig.~\ref{fig:inf_vs_fin}.

The result of the exact calculation for a track of length $L=1.2$
km is also shown in Figs.~\ref{fig:inf_vs_finzhs} and \ref{fig:inf_vs_fin}. 
The agreement between both calculations is excellent because the finite track is
long compared to observation distance. This confirms that 
Eqs.~(\ref{eq:rhofield}) and (\ref{eq:zfield}) must also account for 
Cherenkov radiation. As can be appreciated in Fig.~\ref{fig:inf_vs_fin}, 
the exact results for finite and infinite tracks differ 
for wavelengths larger than the length of the track - frequencies typically
below $\nu_0 \sim (c/n)/\lambda$ with $\lambda \sim L = 1.2$ which gives $\nu \lesim 0.1$ MHz.  

\section{Comparison of the exact calculation and the ZHS algorithm}

We have numerically evaluated the exact expressions for the $z$ and $\rho$ components
of the electric field in Eqs.~(\ref{eq:rhofield}) and (\ref{eq:zfield}) at different frequencies
and observer distances, for a single tracks of different lenghts, and for the tracks
constituting a shower in ice as obtained in full simulations 
performed with the ZHS Monte Carlo code \cite{ZHS92}. 
In this section we present the results of this comparison. 

\subsection{Fourier components of the electric field for a single track}

The applicability of the ZHS expressions in Eqs.~(\ref{eq:rhofield_ZHS}) and (\ref{eq:zfield_ZHS})
relies on the conditions $1-3$ in Section~\ref{sec:ZHS}.
Eqs.~(\ref{eq:cond3}) and (\ref{eq:cond4}) are easily fulfilled for any wavenumber $k$
simply by dividing the particle track in a sufficiently large number of subtracks. 
Once this is guaranteed, the ZHS expression is applied to every sub-track and the electric field is
obtained adding the corresponding contributions. 
The validity of this procedure will be numerically confirmed below when comparing the exact calculation with the 
electric field obtained using the ZHS algorithm in the manner just described. 
However the condition $kr\gg1$, Eq.~(\ref{eq:cond1}), does not depend on 
the size of the track and cannot be enforced by applying the procedure outlined above. 
As a consequence $kr\gg1$ is an intrinsic limit to the range of observing frequencies and distances 
in which the ZHS algorithm gives accurate results as will be shown in the following.

\begin{figure}[htbp]
\begin{center}
{
\scalebox{0.45}{\includegraphics[angle=0]{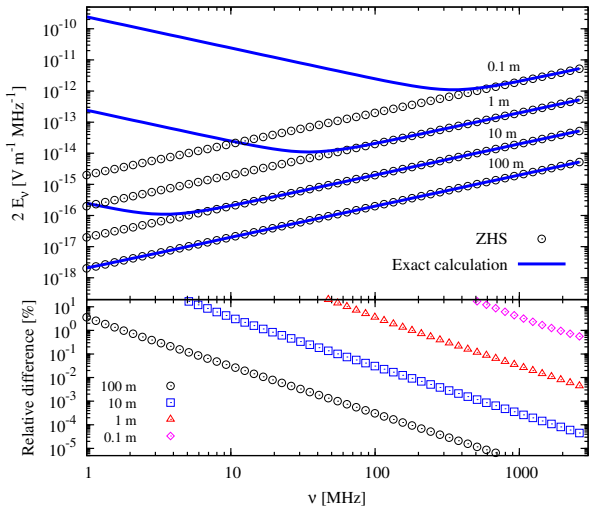}}
}
\caption{
Top panel: Fourier components of the electric field modulus 
for a single particle track as obtained with the exact calculation Eqs.~(\ref{eq:rhofield}) and (\ref{eq:zfield}) (solid lines)
and with the ZHS algorithm Eqs.~(\ref{eq:rhofield_ZHS}) and (\ref{eq:zfield_ZHS}) (open symbols). The length of the track is $L=1.2~10^{-3}$ m 
and the field is shown for observers at distances 
(from top to bottom lines) $R=0.1$ m, 1 m, 10 m and 100 m, 
with respect to the center of the track, and placed at the Cherenkov angle.
Bottom panel: Relative difference (in $\%$) between the exact solution 
and that obtained with the ZHS algorithm for the same distances. 
} 
\label{fig:single_track_1mm}
\end{center}
\end{figure}

In Fig.~\ref{fig:single_track_1mm} we compare the Fourier components of the electric field modulus 
for a single particle track as obtained with the exact calculation, Eqs.~(\ref{eq:rhofield}) and (\ref{eq:zfield}), 
to that obtained with the ZHS algorithm. The length of the track is 
chosen to be small $L=1.2~10^{-3}$m (close to
the peak value of the distribution of track
lengths in the standard ZHS code). 
To test the validity of the ZHS algorithm, we have calculated 
the spectra for observers at different distances ($R$) measured with respect to the center of the track
and placed at the Cherenkov angle. 

The condition $kR\gg1$ in ice with refractive index $n=1.78$ can be cast as:
\begin{equation}
kR \sim 3.7~\left(\frac{\nu}{100~{\rm MHz}} \right)~\left(\frac{R}{1~{\rm m}} \right) \gg 1
\label{eq:kR_ice}
\end{equation}
For observers at distances $R=100,~10,~1$ and $0.1$ m from the particle track, the condition in Eq.~(\ref{eq:kR_ice}) 
is fulfilled as long as $\nu\gesim 1, 10, 100$ MHz and 1 GHz
respectively. The ZHS algorithm is expected to reproduce the results
of the exact calculation in this range. This can be seen 
in Fig.~\ref{fig:single_track_1mm}. The relative difference between
the ZHS and exact calculations is less than $\sim 2\%$ at the frontier
of the validity range in the explored frequency and distance space. The accuracy
can be however orders of magnitude better. If $kR>37$ is enforced for instance 
the corresponding relative difference is below $\sim 0.01\%$. 
This condition is satisfied for $R>10$ m and $\nu\gesim 100$ MHz what 
corresponds to a range of frequencies and distances typically encountered 
in experiments that search for neutrino induced radio transients.  

\begin{figure}[htbp]
\begin{center}
{
\scalebox{0.35}{\includegraphics[angle=-90]{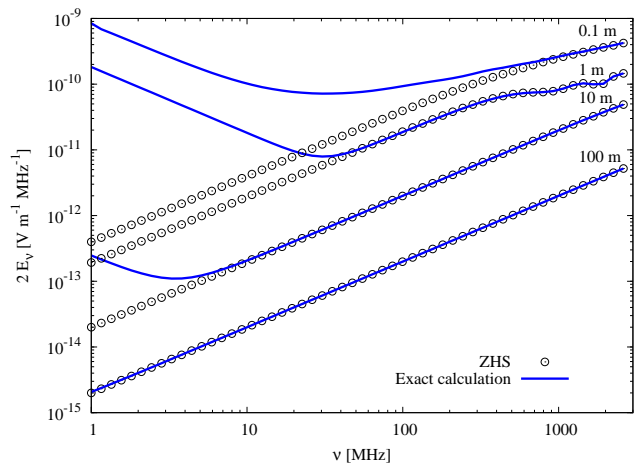}}
}
\caption{
Same as top panel of Fig.~\ref{fig:single_track_1mm} for a single particle track of length $L=1.2$ m. 
} 
\label{fig:single_track_1m}
\end{center}
\end{figure}

What is striking is that these conclusions also apply in the ``near
field'' provided the track is subdivided in sufficiently small sub-tracks. This can be seen in 
Fig.~\ref{fig:single_track_1m} comparing the exact and ZHS results for
a $L=1.2$~m track. The same range of validity is obtained because 
conditions~(\ref{eq:fraunhofer}), (\ref{eq:cond3}) and (\ref{eq:cond4})  
are guaranteed by reducing the length of the sub-tracks. 
%

\begin{figure}[htbp]
\begin{center}
{
\scalebox{0.45}{\includegraphics[angle=0]{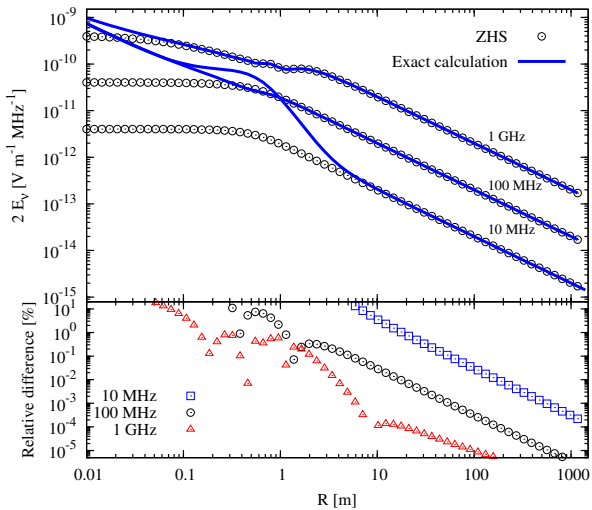}} 
}
\caption{
Top panel: Fourier components of the electric field modulus 
at $\nu=1$ GHz, 100 MHz and 10 MHz (from top to bottom lines)
for a single particle track as a function of distance to the track,
as obtained with the exact calculation Eqs.~(\ref{eq:rhofield}) and (\ref{eq:zfield}) (solid lines)
and with the ZHS algorithm Eqs.~(\ref{eq:rhofield_ZHS}) and (\ref{eq:zfield_ZHS}) (open symbols). The length of the track is $L=1.2$ m 
and the observers are placed at the Cherenkov angle.
Bottom panel: Relative difference (in $\%$) between the exact solution 
and that obtained with the ZHS algorithm for the same frequencies. 
} 
\label{fig:single_track_distance}
\end{center}
\end{figure}

For fixed frequencies the condition in Eq.~(\ref{eq:kR_ice}) only applies at sufficiently large distances $R$
to the track. This is illustrated in Fig.~\ref{fig:single_track_distance} for a track of length $L=1.2$~m. 
The Fourier components at frequencies of $\nu=10$, 100  MHz and 1~GHz
are in agreement with the exact calculation respectively at $R\gesim 10$, 1 and
0.1~m as expected. Since the typical distance between antennas in experiments
such as the Askaryan Radio Array (ARA) \cite{ARA} is $\sim 10 - 100$ m,
we expect the results of the ZHS algorithm obtained through the procedure outlined above, 
to be accurate enough in most practical situations.  

It has been questioned whether the ZHS algorithm reproduces Cherenkov
radiation from a single charged particle track~\cite{endpoints}.
In Figs.~\ref{fig:inf_vs_finzhs} and \ref{fig:inf_vs_fin} the
algorithm is shown to be in very good agreement with the exact solution for 
a 1.2 km track as long as $kR\gg1$ as explained above. The emission
from such track is in turn practically equivalent to the Cherenkov emission from an infinite
track (also displayed) and it follows that the ZHS calculation must
account for Cherenkov radiation. 
%

\subsubsection{Behaviour of the field with frequency}

As can be seen in Fig.~\ref{fig:single_track_1mm} (solid lines) the exact solution for 
the modulus of the electric field scales linearly with frequency provided that $kR\gg 1$. 
As a result the electric field can be approximated with the ZHS expressions, 
Eqs.~(\ref{eq:rhofield_ZHS}) and (\ref{eq:zfield_ZHS}).
For an observer close to the Cherenkov angle as in Fig.~\ref{fig:single_track_1mm}, 
the factor $(\omega-\mathbf{k}\cdot\mathbf{v})\ll 1$ 
and the term in brackets in Eqs.~(\ref{eq:rhofield_ZHS}) and (\ref{eq:zfield_ZHS}) can 
be approximated by: 
\begin{equation}
\begin{split}
&
~\left[ \frac{e^{i(\omega-\mathbf{k}\cdot\mathbf{v})t_2}~
-e^{i(\omega-\mathbf{k}\cdot\mathbf{v})t_1}}{i(\omega-\mathbf{k}\cdot\mathbf{v})} \right]
\approx
\\
& 
\frac{1 + i(\omega-\mathbf{k}\cdot\mathbf{v})t_2 - 1 - i(\omega-\mathbf{k}\cdot\mathbf{v})t_1}{i(\omega-\mathbf{k}\cdot\mathbf{v})} 
= t_2 - t_1 
\end{split}
\end{equation}
making the dependence of the field with $\omega$ apparent.
Clearly the field cannot grow indefinetely with frequency. This 
apparent ``ultraviolet divergence" 
is only an artifact of considering the unrealistic 
medium in which the permittivity $\epsilon(\omega)$ is constant with frequency. 
In a physical medium absorption at high frequencies will tame the growth with frequency 
of the electric field.  

When $kR<1$ the field behaves with frequency as $\omega^{-1}$  as can 
be also seen in Fig.~\ref{fig:single_track_1mm}.
In the model of a charged particle at rest for $t\leq t_1$, 
moving with a speed $v$ between $t=t_1$ and $t=t_2$, 
and becoming again at rest for $t\geq t_2$,
the Coulomb field dominates at small distances to the track and/or 
low frequencies. The field can be modeled as:
\begin{equation}
E_1 \propto \frac{1}{r_1^2}~\Theta(t_{1,obs}-t)
\end{equation} 
for $t<t_1$ and 
\begin{equation}
E_2 \propto \frac{1}{r_2^2}~\Theta(t-t_{2,obs})
\end{equation}
for $t>t_2$, where $r_1$ and $r_2$ are respectively the distances from
the charge to the observer at times $t_1$ and $t_2$, and 
$t_{1,obs}$ and $t_{2,obs}$ denote the instants of time 
at which the Coulomb field arrives at the observer. 
The Fourier transform of a Heaviside function
at non-zero frequency is proportional to $\omega^{-1}$, and 
the two Coulomb fields interfere coherently at low frequencies,
explaining the frequency dependence of the electric field.
Naturally the ZHS algorithm does not reproduce this behavior which is
not associated to radiation.  The growth of the field at low frequencies is 
an artifact of not accounting for screening of the field by the atoms in the medium. 

\subsubsection{Behavior of the field with distance}

In Fig.~\ref{fig:single_track_distance} the dependence of the Fourier components of the field modulus 
with distance is shown for several frequencies. At sufficiently large distances to the track the electric field behaves 
as $1/R$ for all frequencies. This is the radiation zone, the field is expected to behave as $1/R$ as explained 
in conventional radiation theory \cite{Jackson}.
If the observer is placed at very small distances compared to the length of the track, 
the situation resembles that of an infinite track. In this case the discussion 
in Section~\ref{sec:Cherenkov} applies. The field behaves as 
$1/\sqrt{R}$ at distances much smaller than the length of the track provided the
frequency is high enough to satisfy $|u\rho|\gg 1$ - see
Eq.~(\ref{eq:u}). This can be seen in the curve of the Fourier
component at $\nu=1$ GHz for $R \lesim 0.5$~m. 
At small distances and sufficiently low frequencies, when $|u\rho|\ll 1$, the field becomes proportional 
to $1/R$ and independent of frequency - see Eq.~(\ref{eq:zfield_inf}).
This feature can be appreciated in
Fig.~\ref{fig:single_track_distance} at distances below 0.1~m for the
calculations at 10 and 100 MHz. The ZHS algorithm reproduces the
calculation provided the $kR\gg1$ condition is satisfied as could be
expected, reproducing both the $1/R$ and $1/\sqrt{R}$ behaviors. 

\subsection{Fourier components of the electric field in electromagnetic showers}

It is possible to test the ZHS algorithm in a more realistic
situation. In this section we compare the Fourier components of the
electric field predicted by the ZHS algorithm with those obtained 
using the exact calculation in a full simulation 
of electromagnetic showers. 

For this purpose we have applied the exact solutions of the field 
of a track given in Eqs.~(\ref{eq:rhofield}) and (\ref{eq:zfield})
in the ZHS Monte Carlo code \cite{ZHS92} for the simulation of 
electron and photon-induced showers in ice. The ZHS Monte Carlo  
calculates the start and end points of small sub-tracks of all  
charged particles (electrons and positrons) in an electromagnetic 
shower down to a kinetic energy threshold of $\sim 100$ keV. 
With these we can calculate the exact 
electric field produced by each single sub-track and add the fields 
up accounting for interference between different tracks.
Since  Eqs.~(\ref{eq:rhofield}) and (\ref{eq:zfield}) 
are only valid for a charged particle travelling 
along the $z$ axis (parallel to the shower axis), 
we perform the necessary rotations of Eqs.~(\ref{eq:rhofield}) 
and (\ref{eq:zfield}) to obtain the field for a particle
track moving along an arbitrary direction.

Simultaneously with the exact calculation, we also obtain the field 
as predicted by the ZHS algorithm for exactly the same shower 
(i.e. the same set of tracks and sub-tracks). As explained above the subdivisions are 
such that the conditions in Eqs.~(\ref{eq:fraunhofer}), (\ref{eq:cond3}) and 
(\ref{eq:cond4}) are fulfilled for all the sub-tracks in the shower. 

\begin{figure}[htbp]
\begin{center}
{
\scalebox{0.30}{\includegraphics[angle=-90]{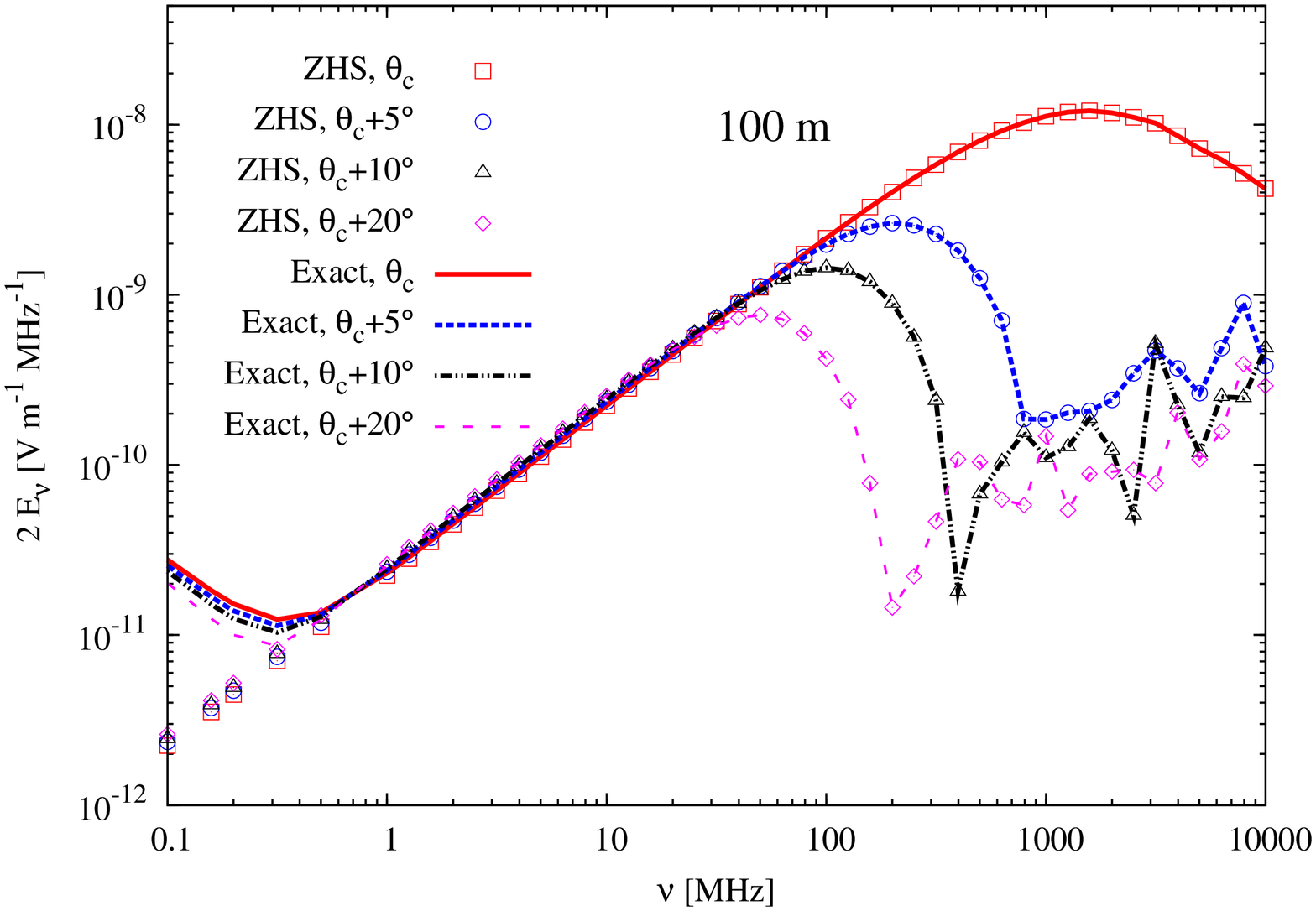}}
\scalebox{0.415}{\includegraphics[angle=0]{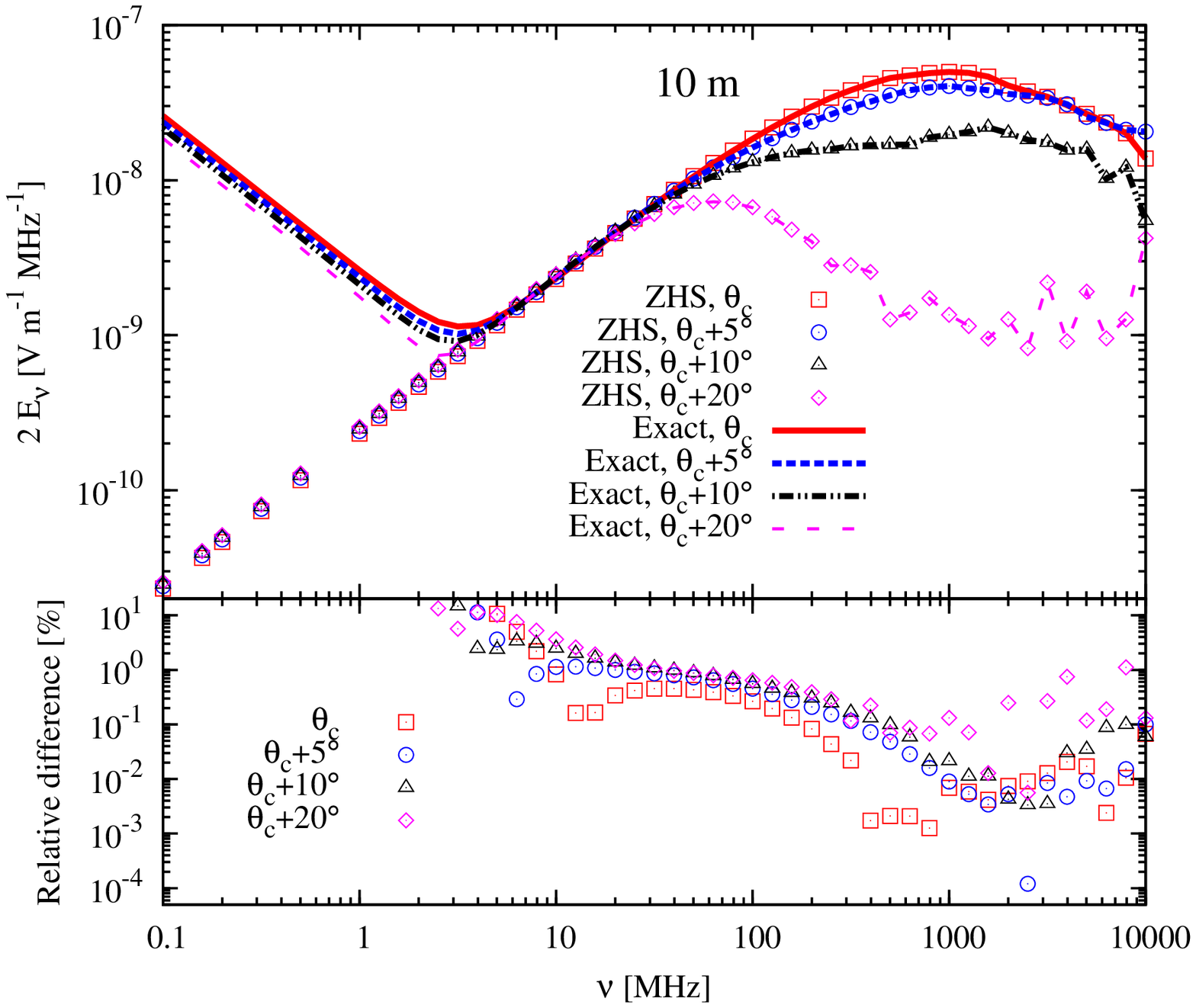}}
\scalebox{0.30}{\includegraphics[angle=-90]{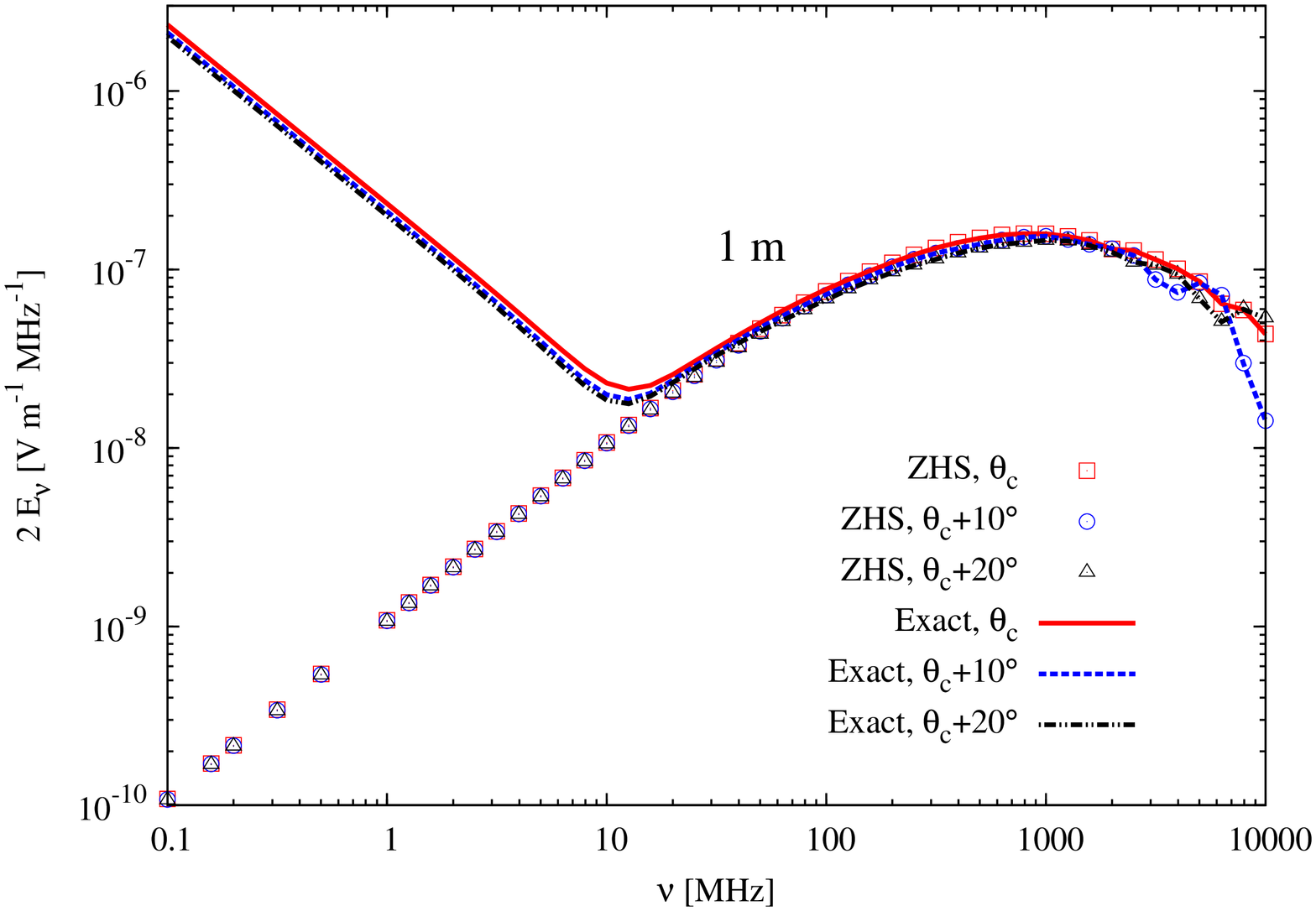}}
}
\caption{
Fourier components of the electric field modulus 
as obtained in Monte Carlo simulations of a 10 TeV electron-induced shower in ice, 
with the exact calculation (lines) and with the ZHS algorithm (symbols). 
The field is shown for observers at distances (from top to bottom panels) 
$R=100$, 10, 1 m, placed at different observation angles with respect to shower maximum.
In the middle panel corresponding to $R=10$ m, we also show the relative difference (in $\%$) between the electric field
modulus as obtained with the exact solution 
and with the ZHS algorithm for the various observation angles depicted.
} 
\label{fig:ZHS_10TeV}
\end{center}
\end{figure}

The result is qualitatively the same as in the case of single tracks. 
As long as the condition $kR\gg1$ is fulfilled, the ZHS algorithm gives an 
accurate prediction for the Fourier components of the electric field
with a difference of less than a few percent relative to those obtained with the exact calculation. 
As can be seen in Fig.~\ref{fig:ZHS_10TeV} 
this occurs for distances to the shower axis as small as $R=1$ m and frequencies
above $\nu\sim10$ MHz, well in the distance and frequency ranges relevant
for experiments looking for particle shower induced radio pulses in
dense media \cite{ARA,ARIANNA,RICE03,ANITA_2009_limits}.

We stress here that the accuracy reported above refers to the
approximation of using the ZHS formula applied to the standard
subdivision of tracks in the ZHS code, instead of the exact expression for 
the radiation emitted by the same particle sub-tracks. 
By comparing the results obtained in the Fraunhofer limit
with the standard subdivision of tracks to those obtained with a much
finer subdivison, it was determined that the accuracy of the 
ZHS code is $\sim 10\%$ at frequencies $\sim 5$~GHz,
improving significantly at lower frequencies.
We do not further address this uncertainty in this paper, 
nor the uncertainty due to the shower simulation itself.

It is also worth remarking that in terms of computing time the exact calculation 
is roughly a factor $\sim 5$ slower than the calcuation performed with 
the ZHS algorithm.

Since the ZHS algorithm can only be applied in a limited range of frequencies, 
an accurate representation of the electric field in the time-domain cannot be obtained
with an inverse Fourier transform of Eqs.~(\ref{eq:rhofield_ZHS}) and (\ref{eq:zfield_ZHS}). 
The low frequency components which do not satisfy the condition
$kR\gg1$ are not accurately described by the ZHS algorithm as shown before. 
Also at very high frequencies the number of steps in which the tracks have to be divided 
in order to fulfill Eqs.~(\ref{eq:fraunhofer}), (\ref{eq:cond3}) and (\ref{eq:cond4}) 
can become prohibitively large from the computational point of view
what can compromise the calculation of the pulses below the 100 picosecond scale. 
In practice however these techniques, although not exact, provide fast and accurate
calculations in the region of interest to UHE neutrino detection. 
For experiments with typical time resolutions of the order of 1 ns, and which 
are only sensitivity to frequencies from $> 10$ MHz to few GHz, the ZHS algorithm 
has been shown to give a very accurate representation of the Fourier components 
of the electric field (Fig.~\ref{fig:ZHS_10TeV}).  

\section{Comparison to other calculations}

Several calculations of the field emitted in showers developing in dense media 
can be found in the literature. 
In \cite{Chih10} the Finite Diference Time Domain method is used for calculating
the field of a pancake-like shower with a Gaussian longitudinal development and Gaussian
radial profile in the time-domain which is then transformed to the frequency-domain. 
In~\cite{buniy02} using the saddle-point approximation, an analytic equation for 
the calculation of the electric field of a charge distribution exhibiting a longitudinal 
profile with a well-pronounced maximum is derived. The result is factorized into 
an integral accounting for the longitudinal variation of the charge and a form 
factor that accounts for the lateral spread of the shower, a procedure revisited in \cite{ARZ11}
for realistic showers. Assuming a Gaussian
longitudinal and lateral development for the charge distribution both results 
were directly compared and turned out to be in good overall agreement as shown in \cite{Chih10}.   
Minor differences could be attributed to the form factors used.

With the exact calculation of the electric field performed in this work, the field due to 
a Gaussian profile can also be obtained and compared to the calculations mentioned
before. The electric current for a shower with a Gaussian profile 
is given by Eq.~(\ref{eq:current}) with the following replacements:
\begin{equation}
Z(t) = 1
\end{equation}
and
\begin{equation}
\label{eq:current_Gaussian}
P(\mathbf{x},t) = \frac{N}{2\pi \sigma_r^2}~e^{-(x^2+y^2)/2\sigma_r^2}~e^{-z^2/2\sigma_l^2}
\delta(z-vt)
\end{equation}
The shower develops in the longitudinal direction parallel to the $z'$ coordinate (shower axis), 
and radially along the $x'$ and $y'$ coordinates. $N$ is a normalization constant. $\sigma_l$ characterizes the 
width of the shower along shower axis and $\sigma_r$ the corresponding lateral width. 
After substituting this current in Eq.~(\ref{eq:vectorpotential}), and following the same steps 
as in Section~\ref{sec:exact}, the expression for the field is the same
as in Eqs.~(\ref{eq:rhofield}) and~(\ref{eq:zfield}) with the following change:
\begin{equation}
\label{eq:field_Gaussian}
\int \mathrm{d} t' \ \ \rightarrow \ \
\int \mathrm{d} t' \ \mathrm{d} x' \ \mathrm{d} y'
\frac{N}{2\pi \sigma_r^2}~e^{-(x'^2+y'^2)/2\sigma_r^2}~e^{-v^2 t'^2/2\sigma_l^2}
\end{equation}
Eqs.~(\ref{eq:rhofield}) and (\ref{eq:zfield}) after the changes in Eq.~(\ref{eq:field_Gaussian}) can be solved numerically. 
The ZHS algorithm can also be applied to this situation as long as the condition $kr\gg1$ for all distances $r$ to 
the shower. The procedure consists on slicing the volume occupied by
the bulk of the shower in small cubes and  
approximating each as a track with constant charge given by the Gaussian distributions in Eq.~(\ref{eq:current_Gaussian}). 

\begin{figure}[htbp]
\begin{center}
{
\scalebox{0.35}{\includegraphics[angle=-90]{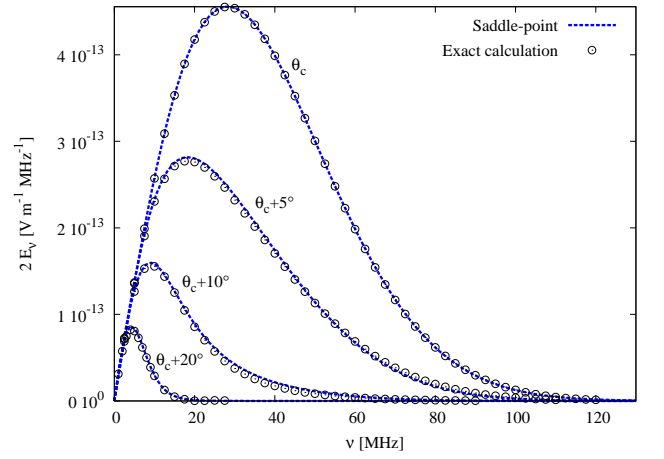}}
}
\caption{
Fourier components of the modulus of the electric field for a Gaussian charge profile 
given in Eq.~(\ref{eq:current_Gaussian}) with
$\sigma_l = 20$ m and $\sigma_r = 1$ m for an observer at $R = 300$ m with respect to
the peak of the Gaussian longitudinal profile. The observation angles are,
from top to bottom, $\theta_C$, $\theta_C + 5^\circ$, $\theta_C + 10^\circ$, and
$\theta_C + 20^\circ$. Fields are calculated with the saddle-point approach and the
exact formula Eqs.~(\ref{eq:rhofield}) and (\ref{eq:zfield}). The result obtained 
with the ZHS algorithm is on top of the exact calculation and it is not plotted in the Fig. for clarity.
} 
\label{fig:comparison}
\end{center}
\end{figure}

Comparison of the result of the exact calculation in this work (or the ZHS algorithm) 
with the saddle-point calculation in \cite{buniy02} requires knowing the form factor $F$ 
for a Gaussian profile. $F$ is defined in \cite{buniy02} as:
\begin{equation}
\label{eq:form_factor}
F(\mathbf{q}) = \int\mathrm{d}x'\mathrm{d}y'\mathrm{d}s' \ e^{-i\mathbf{q}\cdot\mathbf{x'}} f(s',x',y')  
\end{equation}
with $\mathbf{q} = (\omega/v, k\mathbf{\rho}/R)$ with $\mathbf{\rho}=(x,y)$ the radial 
position of the observer. Also, $\mathbf{x}' = (s', x', y')$ with $s' = z' - vt'$.
$R$ is the distance from the maximum of the
shower to the observer.
The function $f$ represents a normalized charge density of the
travelling pancake. Assuming a Gaussian for $f$ of the form: 
\begin{equation}
f(s',x',y') = \frac{1}{2\pi \sigma_r^2} \delta(s')~e^{-(x'^2+y'^2)/2\sigma_r^2},
\end{equation}
and substituting $f$ into Eq.~(\ref{eq:form_factor}), the form factor reads, 
\begin{equation}
F(\mathbf{q})=e^{-\frac{1}{2} (\frac{n\omega}{c}\frac{\rho}{R} \sigma_r)^2}
\end{equation}

Setting $N = 1$, $\sigma_l = 20$ m, $\sigma_r = 1$ m and $R = 300$ m  with
the refractive index of ice $n=1.78$, 
the electric fields for a Gaussian charge profile 
were calculated using the saddle-point approach as described in \cite{buniy02},
and compared to the exact formula given in this work. 
The results are shown in Fig.~\ref{fig:comparison}. Since $R$ is large the condition 
$kR\gg1$ is satisfied for frequencies above $\nu\sim 1$ MHz (see Eq.~\ref{eq:kR_ice}) and the agreement between 
the ZHS and the exact calculation presented in this work is very good (not shown in Fig.~\ref{fig:comparison} 
for clarity). The saddle-point approach 
is also in very good agreement with both the exact and ZHS calculations. 
The results of the field for a Gaussian charge profile are also in good agreement with those obtained in~\cite{Chih10}
using the Finite Difference Time Domain method.

\section{Summary and conclusions}

We have obtained the results of an exact calculation of the electric
field produced by a charged particle that accelerates instantaneously 
moves at constant speed and intantaneously deccelerates. 
The results are used to obtain the approximate expression used in the
ZHS algorithm to calculate the 
radio emission from showers in dense media using shower simulations. 
This allows the precise determination of the conditions necessary for
this approximation to be valid, namely, the observer must be in the far
field zone $kR\gg1$ and the Frauhofer approximation must apply, 
$kL^2 \sin^2 \theta/R\ll1$ for each track. The exact and simulated results are
compared for tracks in a variety of circumstances to illustrate both 
the behavior of the fields and the validity of the approximation made 
in different frequency and distance ranges. 

We have shown that the range of validity of the expressions
can be greatly enlarged by subdividing long charged particle 
tracks in smaller sub-tracks and adding up the conributions, as it is done in
the ZHS algorithm. Nevertheless the far field condition
$kR\gg1$ has been shown to be an intrinsic limit of the ZHS algorithm. 
The ZHS algorithm has been tested for long tracks in regions where the
Fraunhofer approximation does not hold comparing it to the exact
solution. The results clearly indicate that the ZHS algorithm
reproduces the exact behavior provided the far field condition is
satisfied and the length of the sub-tracks is small enough for the
Fraunhofer approximation to be valid. By comparing the emission with
the exact emission from an infinite track it is shown that the emission
calculated with the ZHS algorithm does indeed contain what is
conventionally described as Cherenkov radiation. 
The precision of the ZHS algorithm
is shown to be below the $2\%$ level provided that $kR>3.7$,
corresponding to $R>10$~m and $\nu>10$~MHz. By enforcing $kR>37$ 
the precision improves to better than $0.01\%$, what corresponds to
$R>10$~m and $\nu>100$~MHz, which are conditions met in most
experimental arrangements trying to detect radio pulses form neutrinos
interacting in dense media. 

The ZHS algorithm is tested for completitude when applied to a
shower simulation. This is done comparing the result of the ZHS
algorithm to that obtained when the ZHS expression is replaced by the
exact calculation for every charged particle sub-track for exactly the
same shower. The results
indeed confirm that the accuracies reported for individual sub-tracks
are approximately mantained in the final ZHS result. 

Finally, in order to compare to alternative calculations the
results are compared to the saddle point approximation. This
approximation has been used to test solutions using the method of Finite
Differences to solve Maxwell's Equations directly in the Time Domain (FDTD)
using a simplified shower front based on gaussian distributions in the
shower plane and in time. 
The comparison of the saddle point approximation to both the exact solution
and the ZHS algorithm give compatible results confirming that both
approaches reproduce the radiation emitted from showers. 

The results presented here in summary confirm that the ZHS algorithm
can be used to describe most practical applications to detect pulses emitted
from high energy showers produced in dense media by neutrinos. 
The approach only begins to show significant discrepancies when the observer
is at distances comparable to the lateral dimensions of the shower
($\lesssim$~1~m in ice). Since the typical distance between antennas in experiments
such as the Askaryan Radio Array (ARA) \cite{ARA} is $\sim 10 - 100$ m,
we expect the results to be accurate enough in most practical situations. 

The results obtained are also of interest for the application of the ZHS
algorithm to calculate pulses from EAS. Indeed the application of the
method of track subdivision has been applied in ZHAireS \cite{ZHAireS_air}, 
a recent code developed for this purpose, by making track subdivisions that are
forced to satisfy the Fraunhofer condition.   

\section{Acknowledgments}
J.A-M, W.R.C., D.G.-F. and E.Z. thank Xunta de Galicia
(INCITE09 206 336 PR) and Conseller\'\i a de Educaci\'on
(Grupos de Referencia Competitivos – Consolider Xunta
de Galicia 2006/51); Ministerio de Educaci\'on, Cultura
y Deporte (FPA 2010-18410, FPA2012-39489 and Consolider CPAN -
Ingenio 2010); ASPERA (PRI-PIMASP-2011-1154) and
Feder Funds, Spain. 
We thank CESGA (Centro de SuperComputaci\'on de Galicia) for computing resources.
Part of this research was carried out at the Jet Propulsion
Laboratory, California Institute of Technology, under a
contract with the National Aeronautics and Space Administration.


\end{document}